\documentclass[12pt]{article}

\usepackage[utf8]{inputenc}
\usepackage{textcomp}
\usepackage{amsmath}
\usepackage[shortlabels]{enumitem}
\usepackage{chetnew} 
\usepackage{tikz}
\usepackage{amssymb,amsfonts,mathrsfs,dsfont,yfonts,bbm}\usetikzlibrary{calc}
\usepackage{xcolor}
\usepackage{braket}
\usepackage[normalem]{ulem}
\usepackage{import}
\usepackage{booktabs}
\usepackage{varioref}
\usepackage{bm}
\usepackage{wasysym} 
\usetikzlibrary{arrows.meta,decorations.markings}
\definecolor{slabedge}{HTML}{2B3A55}
\definecolor{slabface}{HTML}{AAC4E2}
\definecolor{surf}{HTML}{12B886}
\definecolor{anyon}{HTML}{E03131}

\newcommand{\beqn}{\begin{eqnarray}}
\newcommand{\eeqn}{\end{eqnarray}}

\newcommand{\be}{\begin{equation}}
\newcommand{\ee}{\end{equation}}
\newcommand{\bea}{\begin{eqnarray}}
\newcommand{\eea}{\end{eqnarray}}

\newcommand{\CD}{\mathcal{D}}

\newcommand{\CC}{\mathcal{C}}

\newcommand{\CO}{\mathcal{O}}

\usepackage{amsmath}	
\usepackage{tikz}
\usepackage{tikz-cd}

\usetikzlibrary{calc,decorations.markings,decorations.pathmorphing,arrows.meta}

\tikzset{
	super thick/.style={line width=3pt}
}
\tikzstyle{knot}=[preaction={super thick, white, draw}]

\usepackage{amsmath}
\usepackage{amsthm}

\theoremstyle{definition}

\newcommand{\doi}[1]{\href{https://dx.doi.org/#1}{{\small DOI:#1}}}

\title{Anyons and Inherently Complex \texorpdfstring{$F$}{F}-symbols}
\author{Matthew Buican,$^{\diamondsuit}$ Peter Huston,$^{\spadesuit}$ and Jiannis K. Pachos$^{\heartsuit}$}

\affiliation{$^{\diamondsuit}$\smallskip Centre for Theoretical Physics and Astronomy\\
Queen Mary University of London, London E1 4NS, UK\\[1mm]$^{\spadesuit}$\smallskip School of Mathematics, University of Leeds, Leeds LS2 9JT, UK \\[1mm] $^{\heartsuit}$\smallskip School of Physics and Astronomy, University of Leeds, Leeds LS2 9JT, UK}

\abstract{Anyons in $2+1$ dimensions are not only characterized by exotic braiding statistics but also by intricate fusion properties. Two anyons may fuse into multiple topological charge sectors, and associativity of fusing three anyons to produce a fixed charge sector is governed by $F$-symbols. While braiding invariants, such as the modular data, are typically complex valued, a complete description of general anyon models requires understanding the arithmetic properties of their fusion associativity data as well. The $F$-symbols for many of the most common $2+1$d topological orders, including all Abelian anyon models as well as Fibonacci and Ising anyons, can be made real valued. We show this phenomenon is not universal by exhibiting unitary premodular categories whose $F$-symbols cannot be made real. We call such $F$-symbols \lq\lq inherently complex." The examples we study lack a charge-conjugation symmetry and our results are therefore consistent with the converse of a statement proved in a companion work linking real $F$-symbols in unitary premodular fusion categories with the existence of a suitable charge-conjugation symmetry. We analyse the smallest-rank unitary premodular fusion categories we know of with inherently complex $F$-symbols: ${\rm Rep}(\mathbb{Z}_7\rtimes\mathbb{Z}_3)$ and ${\rm Rep}(\mathbb{Z}_5\rtimes\mathbb{Z}_4)$. Consequently, the corresponding $\mathcal Z({\rm Rep}(\mathbb{Z}_7\rtimes\mathbb{Z}_3))$ and $\mathcal Z({\rm Rep}(\mathbb{Z}_5\rtimes\mathbb{Z}_4))$ anyon models also have inherently complex $F$-symbols. Our presentation connects these examples with recent results on classifying anyons beyond modular data.}

\begin{document}
\setcounter{tocdepth}{2}
\maketitle
\toc

\section{Introduction}
Anyons arise in a wide range of physical and computational settings, from fractional quantum Hall systems to quantum information platforms that simulate topological matter. Algebraically, anyons correspond to objects in unitary braided spherical fusion categories, {i.e.} unitary premodular categories, with unitary modular tensor categories (UMTCs) describing cases exhibiting non-degenerate braiding.\footnote{Physically, we can think of the degenerate case as corresponding to lines in a topological subsector of a larger, potentially non-topological, quantum field theory.} Fusion specifies the possible topological charge sectors obtained when anyons are combined, while braiding describes the effect of exchanging them. A standard first step in classifying anyon models is therefore to study their modular data, which is generated by the modular $S$ and $T$ matrices \cite{ng2023classification}. However, modular data alone does not distinguish all UMTCs \cite{mignard2021modular}. More generally, recent results indicate that even the full braiding data, encoded in the $R$-symbols, is insufficient to determine a general anyon theory \cite{ng2024recovering}.  As a result, one must also understand the associativity data of fusion, given by the $F$-symbols, which relate different ways of fusing three anyons to produce the same total topological charge. In the language of high-energy physics, one can think of the $F$-symbols as maps between $s$- and $t$-channel decompositions of the same process.

Here, it is important to understand the fundamental arithmetic properties of the categorical data, in particular which number fields the $F$- and $R$-symbols can be defined over (e.g., see \cite{MR2312110}). A particularly basic question along these lines is whether the data of a given anyon model can be represented over the real numbers or whether non-real phases are unavoidable. Beyond its formal significance, this distinction has practical consequences for anyon tomography and interferometry. For example, real associativity data requires the reconstruction of fewer independent amplitude components than non-real data.

The key difficulty is that neither the $F$-symbols nor the $R$-symbols are themselves basis independent: they transform under unitary changes of basis, usually referred to as \lq\lq gauge transformations," of the fusion spaces associated with trivalent anyon vertices. The meaningful question is therefore not whether a particular presentation of the $F$- and $R$-symbols contains non-real phases, but whether these phases are unavoidably present for any gauge choice. Clearly, these phases must be present if there are gauge-invariant quantities built from the $F$- and $R$-symbols (with real coefficients) that are not real. {\it We shall refer to such necessarily non-real $F$- and $R$-symbols as \lq\lq inherently complex."}

Inherently complex $R$-symbols are familiar. Indeed, consider the modular data, which is gauge invariant and built from the $R$-symbols.\footnote{Strictly speaking, the modular data is built from the $R$-symbols, the quantum dimensions, and the fusion matrices. However, the quantum dimensions are real in a unitary fusion category, and the fusion matrices are integral (and hence real). As a further clarification, note that a choice of pivotal structure is implicit in the quantum dimensions and in the trace used to evaluate $S$ and $T$. However, in a unitary fusion category this structure is the unique spherical one with positive dimensions $d(X)={\rm FPdim}(X)\ge1$ (where FPdim is the Frobenius-Perron dimension), so no non-real phase can enter through it.} Generically, the corresponding $S$ and $T$ matrices contain non-real phases (see \cite{Balasubramanian:2024nei} for a classification of premodular categories with real modular data), and this fact implies that, in such cases, the corresponding $R$-symbols cannot be made real by a change of basis.

The situation for $F$-symbols is more subtle. In many important examples, the associativity data can be represented over the real numbers. For instance, in any Abelian UMTC, or more generally in any Abelian premodular category, there is a gauge in which the $F$-symbols take values in $O(1)=\{\pm1\}$ \cite{Quinn:1998un}. The same is true in many non-Abelian examples of central importance, including Fibonacci and Ising anyons. In these cases, all gauge-invariant quantities constructed purely from the $F$-symbols with real coefficients are real valued.

The main purpose of this paper is to show explicitly that inherently complex $F$-symbols do occur in unitary premodular categories and hence in anyon models. Our approach is to construct gauge-invariant quantities built from $F$-symbols in the rank-five representation categories ${\rm Rep}(\mathbb{Z}_7\rtimes\mathbb{Z}_3)$ and ${\rm Rep}(\mathbb{Z}_5\rtimes\mathbb{Z}_4)$ and to show that these quantities are not real.\footnote{Recall that the rank of a fusion category is the number of simple objects, or, physically, the number of simple topological line types. For ${\rm Rep}(G)$, with $G$ finite, this is the number of irreducible representations of $G$.} Thus, no change of splitting basis can make all the corresponding $F$-symbols real. As far as we are aware, these are the lowest-rank unitary premodular categories currently known to exhibit inherently complex $F$-symbols. Moreover, the same obstruction is inherited by any (twisted) Drinfeld center built from these categories. Indeed, a real gauge for the Drinfeld center would restrict to a real gauge for the representation subcategory, contradicting the gauge-invariant complex phases found below.

Our route to these examples is motivated by a companion work \cite{ToAppear}, where we show that the $F$-symbols of a unitary premodular category can be made real whenever a $\mathbb{Z}_2$ (or trivial) charge conjugation is implemented by a particular type of braided autoequivalence we call \lq\lq flat." In this work we will not dwell on the meaning of \lq\lq flat," but rather we simply note that such a charge conjugation is braided.

The present paper explores the converse direction: categories lacking such a symmetry are natural candidates for inherently complex associativity data. The two examples studied here, ${\rm Rep}(\mathbb{Z}_7\rtimes\mathbb{Z}_3)$ and ${\rm Rep}(\mathbb{Z}_5\rtimes\mathbb{Z}_4)$, both lack a charge-conjugation braided autoequivalence and therefore also lack a flat charge-conjugation symmetry. For finite group representation categories, lacking a braided charge conjugation is equivalent to the underlying group lacking a class-inverting automorphism \cite{MR3421083}, as both $\mathbb{Z}_7\rtimes\mathbb{Z}_3$ and $\mathbb{Z}_5\rtimes\mathbb{Z}_4$ do. With this motivation in hand, we explicitly construct non-real gauge-invariant quantities from the $F$-symbols of our two representation categories and hence show their $F$-symbols are inherently complex.

Let us briefly comment on the minimality of these examples. Within the class of representation categories of finite groups, they are natural first candidates.  Indeed, one can explicitly check that no group of order $|G|\le 21$ other than the examples considered here lacks a class-inverting automorphism. Moreover, a classical theorem of Burnside and Miller \cite{burnside1911theory,miller1910groups} implies that every group of order $|G|>12$ has at least five irreducible representations. Combined with the result of \cite{ToAppear} that all groups $G$ with $|G|<20$ have a flat ($\mathbb{Z}_2$ or trivial) charge conjugation symmetry, this discussion shows that ${\rm Rep}(G)$ categories with four or fewer simple objects admit gauges in which their $F$-symbols are real. Thus, rank five is the first place where our mechanism for inherently complex $F$-symbols can appear for ${\rm Rep}(G)$-type categories.

This evidence does not constitute a classification of all lower-rank unitary premodular categories with inherently complex $F$-symbols. In particular, it does not rule out lower-rank premodular categories that are not group representation categories, nor does it exclude higher-order groups with five irreducible representations and no class-inverting automorphism.\footnote{A systematic check of this latter possibility would require examining groups of order up to $60$ using the bound discussed in \cite{burnside1911theory,miller1910groups}.} Nevertheless, the examples studied below are the simplest ${\rm Rep}(G)$-type premodular categories in which inherently complex $F$-symbols arise.\footnote{Any other potential examples with inherently complex $F$-symbols at rank five would have larger categorical dimensions.}

Although we do not attempt a classification of premodular categories with inherently complex $F$-symbols, the examples studied here suggest a broader class exhibiting this phenomenon. The category ${\rm Rep}(\mathbb{Z}_7\rtimes\mathbb{Z}_3)$ belongs to the family of representation categories
\begin{equation}
    {\rm Rep}(\mathbb{Z}_p\rtimes\mathbb{Z}_q)~,
\end{equation}
where $p\ge 7$ and $q\ge 3$ are odd primes with $q\mid p-1$. For all such $p$ and $q$, the underlying group lacks a class-inverting automorphism, and hence the corresponding representation category lacks a braided charge-conjugation autoequivalence. It is plausible that inherently complex $F$-symbols occur more widely in this family, possibly for all faithful actions of $\mathbb{Z}_q$ where $q>2$.

This observation is suggestive in light of the twisted discrete gauge theories studied in \cite{mignard2021modular}, where distinct modular categories can share the same modular data. Many of the groups appearing in that context are closely related to semidirect products of the type considered here.\footnote{The precise families are not identical: \cite{mignard2021modular} imposes additional restrictions on $p$ and $q$; for example, $p=7$ and $q=3$ is not among their cases.} Both phenomena point to the same general lesson, namely, modular data alone need not capture all the information in a topological phase, and invariants built from $F$-symbols can detect structure potentially invisible to braiding data. It would be interesting to understand whether the failure of modular data to determine a UMTC is systematically related to the presence of inherently complex associativity data.

Our second example, ${\rm Rep}(\mathbb{Z}_5\rtimes\mathbb{Z}_4)$, is a braided near-group fusion category (see the full classification in  \cite{thornton2011braided}). The non-symmetric cases all have gauges with real $F$-symbols \cite{ToAppear}. The situation is different for the symmetric examples, which arise as representation categories
\begin{equation}
    {\rm Rep}(\mathbb{F}_{q'}\rtimes\mathbb{F}_{q'}^{\times})~,
\end{equation}
where $q'=p'^k$, $\mathbb{F}_{q'}$ is the finite field of order $q'$, and $\mathbb{F}_{q'}^{\times}$ is its multiplicative group. The case $q'=5$ gives ${\rm Rep}(\mathbb{Z}_5\rtimes\mathbb{Z}_4)$, the smallest example in this family lacking a charge-conjugation symmetry. This fact makes it a natural source of inherently complex $F$-symbols and suggests that further examples may occur for more general values of $q'$.

The rest of the paper is organized as follows. In Sec.~\ref{BraidedFC}, we review the basic definitions of $F$-symbols, $R$-symbols, and their gauge transformations. In Sec.~\ref{WhyBraiding}, we use Abelian fusion categories to illustrate why inherently complex $F$-symbols are easy to obtain in the absence of braiding, thereby motivating the search for braided examples. In Sec.~\ref{Complex}, we analyse our main examples, ${\rm Rep}(\mathbb{Z}_7\rtimes\mathbb{Z}_3)$ and ${\rm Rep}(\mathbb{Z}_5\rtimes\mathbb{Z}_4)$, and construct explicit gauge-invariant complex quantities from their $F$-symbols. We conclude with open questions and possible generalisations.

\section{Braided Fusion Category Basics}
\label{BraidedFC}

In this section, we review the basic categorical concepts needed in the rest of the paper, such as fusion spaces, $F$-symbols, $R$-symbols, and their gauge transformations. The relevant underlying structures are unitary braided spherical fusion categories.
In the case of non-degenerate braiding, these are unitary modular tensor categories, while in the more general (possibly degenerate) case they are often called unitary pre-modular fusion categories or unitary ribbon fusion categories.
Our review is not intended to be exhaustive, and we refer the reader to Chapter 8 of \cite{etingof2015tensor} for a systematic mathematical treatment. The aim here is instead to fix conventions and to present the ingredients in a form that makes contact with the physical language of anyon fusion and braiding.

As discussed above, fusion and braiding are the basic operations that anyons in $2+1$ dimensions can undergo. Physically, an anyon is often viewed as a topological line operator in a $2+1$-dimensional topological quantum field theory. From the perspective of generalised symmetries, such line operators furnish examples of generalised one-form symmetries \cite{Gaiotto:2014kfa}.\footnote{Our terminology is slightly broader than that of \cite{Gaiotto:2014kfa}: we do not require the corresponding line operators to be invertible.}  More generally, topological line operators can also appear in quantum field theories that are not themselves topological, such as Chern-Simons matter theories.

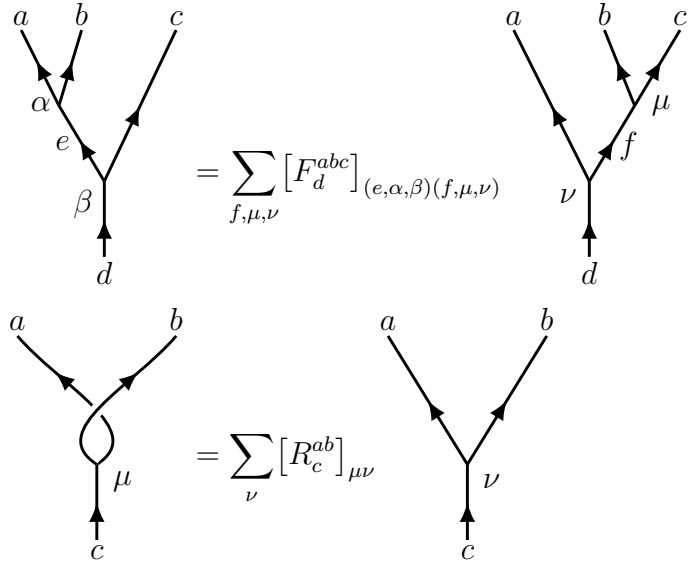
\begin{figure}[!ht]
\begin{align*}
\begin{tikzpicture}[baseline,
    line width=1.1pt,
    >={Latex[length=2.6mm,width=2.1mm]},
    mid/.style={
      postaction={decorate},
      decoration={
        markings,
        mark=at position #1 with {\arrow{Latex[length=2.6mm,width=2.1mm]}}
      }
    },
    mid/.default=0.5,
    every node/.style={inner sep=1.5pt}
  ]
  \begin{scope}[shift={(0,0)}]
    \coordinate (Fd)  at (1.0,-1.0);
    \coordinate (Fb)  at (1.0, 0.0);
    \coordinate (Fa)  at (0.4, 1.0);
    \coordinate (Fla) at (-0.10,2.0);
    \coordinate (Flb) at (0.70, 2.0);
    \coordinate (Flc) at (1.95, 2.0);
    \draw[mid]      (Fd) -- (Fb);
    \draw[mid=0.55] (Fb) -- (Fa); 
    \draw[mid=0.6]  (Fa) -- (Fla);
    \draw[mid=0.6]  (Fa) -- (Flb); 
    \draw[mid]      (Fb) -- (Flc); 
    \node[above] at (Fla) {$a$};
    \node[above] at (Flb) {$b$};
    \node[above] at (Flc) {$c$};
    \node[below] at (Fd)  {$d$};
    \node at (0.16,1.02)  {$\alpha$};
    \node at (0.45,0.50)  {$e$};
    \node at (0.72,-0.30) {$\beta$};
  \end{scope}
  \end{tikzpicture}
  &=
    \displaystyle\sum_{f,\mu,\nu}\bigl[F^{abc}_d\bigr]_{(e,\alpha,\beta)(f,\mu,\nu)}
  \begin{tikzpicture}[baseline,
    line width=1.1pt,
    >={Latex[length=2.6mm,width=2.1mm]},
    mid/.style={
      postaction={decorate},
      decoration={
        markings,
        mark=at position #1 with {\arrow{Latex[length=2.6mm,width=2.1mm]}}
      }
    },
    mid/.default=0.5,
    every node/.style={inner sep=1.5pt}
  ]
    \coordinate (Gd)  at (1.0,-1.0);
    \coordinate (Gn)  at (1.0, 0.0);
    \coordinate (Gm)  at (1.6, 1.0);
    \coordinate (Gla) at (0.00,2.0);
    \coordinate (Glb) at (1.20,2.0);
    \coordinate (Glc) at (2.20,2.0);
    \draw[mid]      (Gd) -- (Gn);
    \draw[mid]      (Gn) -- (Gla);
    \draw[mid=0.55] (Gn) -- (Gm);
    \draw[mid=0.6]  (Gm) -- (Glb);
    \draw[mid=0.6]  (Gm) -- (Glc);
    \node[above] at (Gla) {$a$};
    \node[above] at (Glb) {$b$};
    \node[above] at (Glc) {$c$};
    \node[below] at (Gd)  {$d$};
    \node at (1.95,1.00)  {$\mu$};
    \node at (1.52,0.46)  {$f$};
    \node at (0.72,-0.20) {$\nu$};
  \end{tikzpicture}\\
  \begin{tikzpicture}[baseline,
    line width=1.1pt,
    >={Latex[length=2.6mm,width=2.1mm]},
    mid/.style={
      postaction={decorate},
      decoration={
        markings,
        mark=at position #1 with {\arrow{Latex[length=2.6mm,width=2.1mm]}}
      }
    },
    mid/.default=0.5,
    every node/.style={inner sep=1.5pt}
  ] 
    \coordinate (Rc)  at (1.0,-1.0);
    \coordinate (Rm)  at (1.0, 0.0);
    \coordinate (Rla) at (-0.05,1.7);
    \coordinate (Rlb) at (2.05,1.7);
    \draw[mid=0.7] (Rm) .. controls (1.7,0.45) and (0.5,1.05) .. (Rla);
    \draw[mid=0.7,preaction={draw,line width=5pt,white}]
          (Rm) .. controls (0.3,0.45) and (1.5,1.05) .. (Rlb);
    \draw[mid] (Rc) -- (Rm);
    \node[above] at (Rla) {$a$};
    \node[above] at (Rlb) {$b$};
    \node[below] at (Rc)  {$c$};
    \node at (1.34,-0.22) {$\mu$};
  \end{tikzpicture}
  &=
    \displaystyle\sum_{\nu}\bigl[R^{ab}_c\bigr]_{\mu\nu}
  \begin{tikzpicture}[baseline,
    line width=1.1pt,
    >={Latex[length=2.6mm,width=2.1mm]},
    mid/.style={
      postaction={decorate},
      decoration={
        markings,
        mark=at position #1 with {\arrow{Latex[length=2.6mm,width=2.1mm]}}
      }
    },
    mid/.default=0.5,
    every node/.style={inner sep=1.5pt}
  ] 
    \coordinate (Sc)  at (1.0,-1.0);
    \coordinate (Sn)  at (1.0, 0.0);
    \coordinate (Sla) at (-0.05,1.7);
    \coordinate (Slb) at (2.05,1.7);
    \draw[mid] (Sn) -- (Sla);
    \draw[mid] (Sn) -- (Slb);
    \draw[mid] (Sc) -- (Sn);
    \node[above] at (Sla) {$a$};
    \node[above] at (Slb) {$b$};
    \node[below] at (Sc)  {$c$};
    \node at (1.32,-0.22) {$\nu$};
\end{tikzpicture}
\end{align*}
  \caption{The $F$- and $R$-symbols of a braided fusion category. The $F$-symbols describe the associativity of fusion, while the $R$-symbols describe braiding. Following \cite{Barkeshli:2014cna}, the diagrams are read from bottom to top, and the trivalent vertices represent splitting spaces, $V^{xy}_z\cong{\rm Hom}(z,x\otimes y)$. The arrows record the orientation of the corresponding topological lines. Thus, in the top equation, the left-hand tree is the composite $d\to e\otimes c\to(a\otimes b)\otimes c$, with vertices $\alpha\in V^{ab}_e$ and $\beta\in V^{ec}_d$, while the right-hand tree is $d\to a\otimes f\to a\otimes(b\otimes c)$, with vertices $\mu\in V^{bc}_f$ and $\nu\in V^{af}_d$. Both furnish bases of $V^{abc}_d$, and $\bigl[F^{abc}_d\bigr]$ expresses the left-hand basis in terms of the right-hand one. The $F$- and $R$-symbols depend on the choice of bases for these spaces and therefore transform as in \eqref{eq:gaugeTransformation} and \eqref{RGT}.}
  \label{fig:fr-moves}
\end{figure}

Mathematically, we take topological line operators to correspond to objects in a pre-modular category, $\CC$. Given a set of simple objects, $\left\{\ell_i\right\}$, in $\CC$, we can define fusion via
\begin{equation}\label{fusionL}
\ell_i\otimes\ell_j\cong\ell_j\otimes\ell_i\cong\bigoplus_{\ell_k}N_{\ell_i\ell_j}^{\ell_k}\ell_k~,\ \ \ N_{\ell_i\ell_j}^{\ell_k}\in\mathbb{Z}_{\ge0}~.
\end{equation}
Physically, to each outcome on the RHS, we can assign a three-punctured sphere and a corresponding Hilbert space. Mathematically, this fusion space is referred to as ${\rm Hom}(\ell_i\otimes\ell_j, \ell_k)$, and it has complex dimension $N_{\ell_i\ell_j}^{\ell_k}$ (the adjoint space is referred to as the \lq\lq splitting" space and describes the splitting of objects into products). For any simple object $\ell$, there is a dual object $\ell^{\vee}$ (note that we can have self-dual objects, i.e. $\ell^{\vee}\cong\ell$) such that
\begin{equation}
\ell\otimes\ell^{\vee}\cong1\oplus\cdots~,
\end{equation}
where $1$ is the trivial object,\footnote{That is, $1$ is the identity object for the tensor product, meaning $1\otimes x\cong x\otimes 1\cong x$ for any object $x\in\CC$.} and the ellipses contain potential non-trivial objects. If the ellipses are empty for all simples $\ell\in\CC$, the premodular category is referred to as Abelian, and the fusion rules are those of an Abelian group. In this case, $\ell^{\vee}$ is the inverse of $\ell$. Otherwise $\CC$ is non-Abelian, and the corresponding topological line operators are referred to in the physics literature as constituting a \lq\lq non-invertible" 1-form symmetry.

We can think of \eqref{fusionL} as corresponding to an operator product expansion of topological lines subject to consistency conditions arising from associativity as in the top diagram of Fig. \ref{fig:fr-moves}. In particular, the changes of basis coefficients when one applies associativity are called the $F$-symbols, as shown in Fig. \ref{fig:fr-moves} (throughout, we follow the notation and conventions of \cite{Barkeshli:2014cna}). They satisfy a set of consistency conditions called the pentagon equations
\begin{equation}
\begin{aligned}\label{Pentagon}
&\sum_{\delta}
\left[F^{fcd}_{e}\right]_{(g,\beta,\gamma)(l,\nu,\delta)}
\left[F^{abl}_{e}\right]_{(f,\alpha,\delta)(k,\mu,\lambda)} \\
&\qquad =
\sum_{h,\sigma,\psi,\rho}
\left[F^{abc}_{g}\right]_{(f,\alpha,\beta)(h,\psi,\sigma)}
\left[F^{ahd}_{e}\right]_{(g,\sigma,\gamma)(k,\rho,\lambda)}
\left[F^{bcd}_{k}\right]_{(h,\psi,\rho)(l,\nu,\mu)}~,
\end{aligned}
\end{equation}
which enforce the condition that two different ways of using $F$-moves to reassociate the fusion of four objects produce the same change of basis.
Any solution to these equations defines a valid set of $F$-symbols. Note that these quantities are not unique. Indeed, we have the freedom to perform arbitrary unitary changes of basis, $[\Gamma^{xy}_z]\in U(N_{xy}^z)$, in the four splitting spaces that feed into the definition of the $F$-symbol. Under such changes of bases, the $F$-symbol undergoes the gauge transformation
\begin{eqnarray}
\label{eq:gaugeTransformation}
\big[\widetilde{F}^{abc}_{d}\big]_{(e,\alpha,\beta)(f,\mu,\nu)} 
\!\!&=&\!\! \sum_{\alpha',\beta',\mu',\nu'} 
\left[\Gamma^{ab}_{e}\right]_{\alpha\alpha'}
\left[\Gamma^{ec}_{d}\right]_{\beta\beta'}
\big[F^{abc}_{d}\big]_{(e,\alpha',\beta')(f,\mu',\nu')}\Big[\big(\Gamma^{bc}_{f}\big)^{-1}\Big]_{\mu'\mu}
\Big[\big(\Gamma^{af}_{d}\big)^{-1}\Big]_{\nu'\nu}~.\ \ \ \ \ \ 
\end{eqnarray}

In addition to fusion, objects in a premodular category can undergo braiding, implemented by $R$-symbols, as in the bottom diagram in Fig. \ref{fig:fr-moves}. In order for a category to admit braiding, the $F$-symbols and $R$-symbols must be compatible. This requirement amounts to solving a pair of hexagon consistency equations
\begin{eqnarray}\label{Hexagon}
\sum_{\lambda,\gamma}
\left[R^{ac}_{e}\right]_{\alpha\lambda}
\left[F^{acb}_{d}\right]_{(e,\lambda,\beta)(g,\gamma,\nu)}
\left[R^{bc}_{g}\right]_{\gamma\mu}
\!\!&=&\!\!
\sum_{f,\sigma,\delta,\psi}
\left[F^{cab}_{d}\right]_{(e,\alpha,\beta)(f,\delta,\sigma)}
\big[R^{fc}_{d}\big]_{\sigma\psi}
\left[F^{abc}_{d}\right]_{(f,\delta,\psi)(g,\mu,\nu)},
\nonumber\\
\sum_{\lambda,\gamma}
\Big[\big(R^{ca}_{e}\big)^{-1}\Big]_{\alpha\lambda}
\left[F^{acb}_{d}\right]_{(e,\lambda,\beta)(g,\gamma,\nu)}
\Big[\big(R^{cb}_{g}\big)^{-1}\Big]_{\gamma\mu}
\!\!&=&\!\!
\sum_{f,\sigma,\delta,\psi}
\left[F^{cab}_{d}\right]_{(e,\alpha,\beta)(f,\delta,\sigma)}
\Big[\big(R^{cf}_{d}\big)^{-1}\Big]_{\sigma\psi}
\left[F^{abc}_{d}\right]_{(f,\delta,\psi)(g,\mu,\nu)}
\end{eqnarray}

As in the case of the $F$-symbols, the $R$-symbols are not gauge invariant
\begin{align}\label{RGT}
\big[\widetilde{R}^{ab}_{c}\big]_{\mu\nu}
&= \sum_{\mu',\nu'} 
\left[\Gamma^{ba}_c\right]_{\mu\mu'}
\left[R^{ab}_c\right]_{\mu'\nu'}
\left[(\Gamma^{ab}_c)^{-1}\right]_{\nu'\nu}~,
\end{align}
where $\mu$ and $\nu$ label bases of $V^{ba}_c$ and $V^{ab}_c$, respectively. It is rather straightforward to construct universal invariants from the $R$-symbols. For example, we can construct the generators of the modular group: $T_{aa}=d_a^{-1}\sum_cd_c{\rm Tr}R^{aa}_c$ and $S_{ab}=\frac{1}{\CD}\sum_cd_c{\rm Tr}(R^{b^{\vee}a}_cR^{ab^{\vee}}_c)$, where $c$ runs over simple objects in $\CC$ and $d_i\in\mathbb{R}_{>0}$ is the $S^3$ expectation value of a closed loop of $\ell_i$. Here $T$ is the topological spin, and $S$ is the double braiding (or, more physically, the mutual Aharonov-Bohm phase from winding one topological line around the other). In the case of UMTCs, the $S$ matrix is non-degenerate. Our main examples in Sec. \ref{Complex}, on the other hand, have completely degenerate braiding: the $S$ matrix has rank one. In \lq\lq almost all" pre-modular categories, $S_{ab}$ and $T_{ab}$ are complex \cite{Balasubramanian:2024nei}. This fact implies that in almost all pre-modular categories, $R^{ab}_c$ is inherently complex.

There are also universal invariants that can be expressed in terms of $F$-symbol data, such as Frobenius-Schur indicators \cite{Lake:2016mky}. However, the examples studied in Sec. \ref{Complex} show that these standard invariants do not exhaust the possible gauge-invariant information carried by the $F$-symbols. In particular, for our two main categories of interest, and for their untwisted Drinfeld centers, none of the universal invariants known to us detect the inherent complexity of the associativity data.  We therefore construct more tailored gauge-invariant quantities in Sec. \ref{Complex}, designed specifically to detect complex phases that cannot be removed by a change of splitting basis.

Before concluding this review, let us specify what we mean by a charge-conjugation symmetry in a premodular category. At the level of simple objects, charge conjugation sends each object $\ell\in\CC$ to its dual $\ell^\vee$. This permutation is always visible in the modular data, because in a modular category it is implemented by the matrix $S^2=C$. However, this does not by itself imply that charge conjugation is a genuine symmetry of the category. For our purposes, we require a braided autoequivalence that lifts the object-level assignment $\ell\mapsto\ell^\vee$ to the full categorical data. In particular, it must preserve the $F$- and $R$-symbols up to gauge transformations. In components, this means~\cite{Barkeshli:2014cna}
\begin{equation}
\begin{aligned}
\rho_C\!\left(\left[F^{abc}_{d}\right]_{(e,\alpha,\beta)(f,\mu,\nu)}\right)
&=\sum_{\alpha',\beta',\mu',\nu'}
\left[U_C{}^{a^{\vee}b^{\vee}}_{e^{\vee}}\right]_{\alpha\alpha'}
\left[U_C{}^{e^{\vee}c^{\vee}}_{d^{\vee}}\right]_{\beta\beta'}
\left[F^{a^{\vee}b^{\vee}c^{\vee}}_{d^{\vee}}\right]_{(e^{\vee},\alpha',\beta')(f^{\vee},\mu',\nu')} \\[2pt]
&\quad\times
\left[\left(U_C{}^{b^{\vee}c^{\vee}}_{f^{\vee}}\right)^{-1}\right]_{\mu'\mu}
\left[\left(U_C{}^{a^{\vee}f^{\vee}}_{d^{\vee}}\right)^{-1}\right]_{\nu'\nu} \\[4pt]
&=\left[F^{abc}_{d}\right]_{(e,\alpha,\beta)(f,\mu,\nu)}~,
\end{aligned}
\end{equation}
and
\begin{equation}
\begin{aligned}
\rho_C\!\left(\left[R^{ab}_{c}\right]_{\mu\nu}\right)
&=\sum_{\mu',\nu'}
\left[U_C{}^{b^{\vee}a^{\vee}}_{c^{\vee}}\right]_{\mu\mu'}
\left[R^{a^{\vee}b^{\vee}}_{c^{\vee}}\right]_{\mu'\nu'}
\left[\left(U_C{}^{a^{\vee}b^{\vee}}_{c^{\vee}}\right)^{-1}\right]_{\nu'\nu} \\[4pt]
&=\left[R^{ab}_{c}\right]_{\mu\nu}~.
\end{aligned}
\end{equation}
In the above equations, $\rho_C$ indicates the action of charge conjugation.

\section{Complex \texorpdfstring{$F$}{F}-symbols Without Braiding}
\label{WhyBraiding}

In this section, we step back from braided categories and explain why inherently complex $F$-symbols are easy to obtain when braiding is absent. The examples we discuss here provide useful contrast to the braided examples studied in Sec. \ref{Complex}. In the companion work \cite{ToAppear}, the proof that a suitable $\mathbb{Z}_2$ charge-conjugation autoequivalence leads to real $F$-symbols crucially relies on the presence of braiding.

Here we illustrate the above discussion in the Abelian setting. The results we present in this section essentially follow from \cite{Quinn:1998un} (see also \cite{galindo2025note}), but we include a compact derivation here for completeness and concreteness. As we will show, if an Abelian fusion category has inherently complex $F$-symbols, then it cannot admit a braiding. From the physical point of view, this distinction is important because genuine line operators in $2+1$ dimensions are braided. Thus the non-braided examples discussed below should be viewed instead as fusion categories of non-genuine lines that bound topological surfaces rather than anyon models.

As one of the simplest examples, let us consider the fusion category ${\rm Vec}_{\mathbb{Z}_3}^{[\omega_k]}$ with non-trivial associator 
\begin{equation}
\omega_k(a,b,c):=F^{abc}_{a+b+c}=\exp\left({2\pi ik\over9}a\left(b+c-[b+c]_3\right)\right)~,
\end{equation}
where $[b+c]_3$ means we work mod 3. For $k\ne0$ mod 3, the above expression defines a non-trivial class $[\omega_k]\in H^3(\mathbb{Z}_3,U(1))$. Note that the $F$-symbol is fixed (up to gauge transformations) by the three inputs $a$, $b$, and $c$. Moreover, the pentagon equations \eqref{Pentagon} in this case imply that $\omega_k(a,b,c)$ is a 3-cocycle. It is straightforward to check that, for $k\ne0\ {\rm mod}\ 3$, the above category does not admit a braiding, because it does not satisfy the hexagon equations \eqref{Hexagon}. Importantly, the above $F$-symbol cannot be made real in any gauge, because complex conjugation exchanges $[\omega_1]\leftrightarrow[\omega_2]$, whereas a real $F$-symbol would require the fusion category, and therefore cohomology class, invariant under complex conjugation. By contrast, any braided pointed category with fusion group $\mathbb Z_3$ has trivial ordinary associator class, $[\omega]=[1]$, and hence admits a gauge in which $F^{abc}_{a+b+c}=1$. This reality is consistent with the result in \cite{ToAppear}, because ${\rm Vec}_{\mathbb{Z}_3}^{[\omega_0]}$ has a braided charge conjugation autoequivalence.\footnote{This statement follows from the fact that $q(g^{-1})=q(g)$ for the quadratic form that determines the topological spins.}

This example illustrates how complex associator phases can arise in the absence of braiding. We now generalize this example as follows: we prove that if an Abelian fusion category has inherently complex $F$-symbols, then it does not admit a braiding.

To derive this statement, we prove the contrapositive. To that end, let $\mathcal C$ be a braided Abelian fusion category, with associator $\omega(a,b,c)$ and braiding phases $R^{ac}:=R^{ac}_{a+c}$. In this case, the hexagon equations \eqref{Hexagon} reduce to
\begin{eqnarray}
R^{ac}R^{bc}\cdot{\omega(a,c,b)\over\omega(c,a,b)\omega(a,b,c)}&=&R^{a+b,c}~,\cr R^{ca}R^{cb}\cdot{\omega(c,a,b)\omega(a,b,c)\over\omega(a,c,b)}&=&R^{c,a+b}~.
\end{eqnarray}
Next, note that the $S$-matrix becomes $S^{ab}=R^{ab}R^{ba}$, which is a symmetric bicharacter. Indeed, symmetry is trivial, and the bicharacter property follows from
\begin{eqnarray}\label{symBC}
S^{a,b+c}&=&R^{a,b+c}R^{b+c,a}=\left(R^{ba}R^{ca}\cdot{\omega(b,a,c)\over\omega(a,b,c)\omega(b,c,a)}\right)\left(R^{ab}R^{ac}\cdot{\omega(a,b,c)\omega(b,c,a)\over\omega(b,a,c)}\right)\cr&=& S^{ab}S^{ac}~.
\end{eqnarray} 

With this understanding, we can now extract a useful consequence. Suppose that the pair $(\omega(a,b,c),R^{ab})$ satisfies the Abelian hexagon equations. Then the pair
\begin{equation}
    \big(\omega(a,b,c)^2,(R^{ab})^2\big)~,
\end{equation}
also satisfies the hexagon equations. The topological spin associated with this new braided Abelian structure is
\begin{equation}
    \widetilde{\theta}_a=(R^{aa})^2=\theta_a^2~.
\end{equation}
On the other hand, the pair $(1,S^{ab})$ also satisfies the hexagon equations by \eqref{symBC}. Its topological spin is
\begin{equation}
    S^{aa}=R^{aa}R^{aa}=\theta_a^2~.
\end{equation}
Thus the two braided Abelian structures
\begin{equation}
    \big(\omega^2,R^2\big)
    \qquad\text{and}\qquad
    (1,S)~,
\end{equation}
have the same quadratic form. By the results in \cite{eilenberg1953groups,eilenberg1954groups,street1993braided}, the quadratic form determines the braided Abelian category up to equivalence.  It follows that $\big(\omega^2,R^2\big)$ is gauge equivalent to $(1,S)$.  In particular, $\omega^2$ is cohomologically trivial,
\begin{equation}
    [\omega^2]=[1]~.
\end{equation}
Equivalently,
\begin{equation}
    [\omega]=[\omega^{-1}]=\overline{[\omega]}~.
\end{equation}
Hence the associator class is invariant under complex conjugation, and the $F$-symbols can be chosen to be real. Indeed, such a choice is guaranteed by the fact that the $F$-symbols in this case are numbers and not higher-dimensional matrices.
This conclusion is consistent with the more explicit statement of
\cite{Quinn:1998un}, which constructs a gauge for braided Abelian categories in which all
associators take values in $\{\pm1\}$.  Hence the $F$-symbols of any braided Abelian
fusion category can be chosen to be real.

Taking the contrapositive, we obtain the following useful statement:

\bigskip
\noindent
{\bf Fact.} If an Abelian fusion category has inherently complex $F$-symbols, then it does not admit a braiding.

\bigskip\noindent
This Abelian result illustrates why inherently complex $F$-symbols are relatively easy to find in non-braided fusion categories. The more interesting question, and the one we turn to next, is whether such complexity can occur in genuinely braided fusion categories.

\section{Examples of Inherently Complex \texorpdfstring{$F$}{F}-symbols}
\label{Complex}

We now arrive at our main examples. In the previous section, we argued that it is straightforward to find examples of inherently complex $F$-symbols among fusion categories that do not admit braiding. In this section, we return to focus on premodular (braided) fusion categories. Physically, the distinction is that now we discuss systems of genuine line operators in $2+1$d, whereas in the previous section we studied lines attached to surfaces. As described in the introduction, many of the most common examples of premodular fusion categories admit real $F$-symbols. Indeed, by our discussion in the previous section, all premodular Abelian fusion categories admit real $F$-symbols. Moreover, various common non-Abelian examples like the Fibonacci and Ising anyons also admit real $F$-symbols.

As discussed in the introduction, to arrive at examples with inherently complex $F$-symbols, it is useful to study categories that lack a braided charge conjugation autoequivalence. This statement follows from our upcoming paper \cite{ToAppear}, which guarantees a gauge with real $F$-symbols when the theory has a special type of braided charge conjugation symmetry. In the absence of a suitable charge conjugation symmetry, it is natural to wonder if we have inherently complex $F$-symbols. In this section, we partly answer this question by exhibiting two such examples: ${\rm Rep}(\mathbb{Z}_7\rtimes\mathbb{Z}_3)$ and ${\rm Rep}(\mathbb{Z}_5\rtimes\mathbb{Z}_4)$.

While these two categories are non-modular (their braiding is completely degenerate), any discrete gauge theory built on the above groups, for any Dijkgraaf-Witten twist, will also have non-real $F$-symbols. The reason is that such a theory has the above categories as braided fusion subcategories
\begin{equation}\label{GT}
{\rm Rep}(\mathbb{Z}_7\rtimes\mathbb{Z}_3)<Z^{\omega_1}(\mathbb{Z}_7\rtimes\mathbb{Z}_3) ~, \ \ \ {\rm Rep}(\mathbb{Z}_5\rtimes\mathbb{Z}_4)<Z^{\omega_2}(\mathbb{Z}_5\rtimes\mathbb{Z}_4)~,
\end{equation}
where the $\omega_i$ are arbitrary Dijkgraaf-Witten twists in $H^3(G,U(1))$.\footnote{Physically, this statement holds because we can always prepare such theories by gauging a $G$ 0-form symmetry of a $G$-symmetry-protected topological phase.} In particular, restricting the $F$-symbols to the above representation subcategories is guaranteed to produce non-real numbers. Note that this route is considerably simpler than constructing the full set of $F$-symbols for the discrete gauge theories themselves.

Before moving on to detailed demonstrations of inherently complex $F$-symbols, let us summarize some of the key features of our examples:
\begin{itemize}
\item We are able to rule out braided charge conjugation symmetries in ${\rm Rep}(\mathbb{Z}_7\rtimes\mathbb{Z}_3)$ and ${\rm Rep}(\mathbb{Z}_5\rtimes\mathbb{Z}_4)$ (and therefore also in the gauge theories discussed in \eqref{GT}) using \cite{MR3421083}. In particular, this fact follows from the observation that neither of the corresponding groups has a class-inverting automorphism.
\item The examples we have constructed are the simplest ones we have been able to find, in the sense that they have the lowest rank and the two lowest categorical dimensions: they both have five simple objects and categorical dimensions given by the orders of the corresponding groups (i.e., twenty-one and twenty respectively).\footnote{However, we make no definite claims as to whether there are lower-rank categories exhibiting the phenomenon of inherently complex $F$-symbols.} The ${\rm Rep}(\mathbb{Z}_7\rtimes\mathbb{Z}_3)$ example in Sec. \ref{Z7Z3} is simpler to analyze: our invariant detector of $F$-symbol complexity is a particular gauge-invariant $F$-symbol of rank one
\begin{equation}
F^{\chi\rho\rho}_{\rho}~,
\end{equation}
where the precise details and gauge invariance are described around \eqref{FINV} below. Turning to the case of ${\rm Rep}(\mathbb{Z}_5\rtimes\mathbb{Z}_4)$ in Sec. \ref{Z5Z4}, our gauge-invariant inherent complexity detector is 
\begin{equation}
{\rm Tr}\left(F^{\chi\rho\rho}_{\rho}\left(F^{\rho\rho\chi}_{\rho}\right)^{-1}\right)~,
\end{equation}
where the relevant details are described around \eqref{Fcomplex} below. While this second example has a more complicated non-real invariant, it is built from a smaller group.\footnote{Indeed, one can explicitly check that all other groups of order twenty-one or less have a class-inverting automorphism and hence do not have inherently complex $F$-symbols for their representation categories.}
\item As discussed in the introduction, we include both examples below, because we expect they may potentially exhibit interesting generalizations. For example, we expect ${\rm Rep}(\mathbb{Z}_7\rtimes\mathbb{Z}_3)$ to be part of a larger class of ${\rm Rep}(\mathbb{Z}_p\rtimes\mathbb{Z}_q)$ fusion categories with primes $p$ and $q$ such that $q|p-1$ potentially having inherently complex $F$-symbols. Many of these categories are known to be subcategories of discrete gauge theories exhibiting modular isotopy: distinct MTCs with the same modular data (i.e., $S$ and $T$) \cite{mignard2021modular}. It is interesting to note that both phenomena (complex $F$ and modular isotopy) require understanding topological phases beyond their modular data.

Finally, we also include the ${\rm Rep}(\mathbb{Z}_5\rtimes\mathbb{Z}_4)$ example because, in addition to $\mathbb{Z}_5\rtimes\mathbb{Z}_4$ being the smallest group whose representation category lacks a braided charge conjugation symmetry, we expect ${\rm Rep}(\mathbb{Z}_5\rtimes\mathbb{Z}_4)$ is the simplest of many near-group categories that potentially have inherently complex $F$-symbols. 
\end{itemize}

As a guide for readers not interested in the full technical derivations, we note that the main results are in \eqref{eq:F-result-Z7Z3} and \eqref{complexTrace} respectively.

\subsec{\texorpdfstring{${\rm Rep}(\mathbb{Z}_7\rtimes\mathbb{Z}_3)$}{Rep(Z7 semidirect product  Z3)}}\label{Z7Z3}
In this subsection, we consider what is in many ways the simplest example of a unitary premodular fusion category with inherently complex $F$-symbols: ${\rm Rep}(\mathbb{Z}_7\rtimes\mathbb{Z}_3)$. By our above discussion, any discrete gauge theory built from a $\mathbb{Z}_7\rtimes\mathbb{Z}_3$ gauge group will also have inherently complex $F$. Although we will find a representation category for a smaller group exhibiting this same phenomenon in the next subsection, the example we consider here has a gauge-invariant $F$-symbol that is non-real, and comes from a group of length $2$. At the level of the braided fusion categories themselves, both examples we consider have rank five.

To find inherently complex $F$-symbols, we should avoid categories with braided $\mathbb{Z}_2$ charge conjugation symmetries. To that end, we recall from Theorem 2.14 of \cite{MR3421083} that non-Abelian groups of odd order lack a class-inverting automorphism. The group we are studying here is the smallest odd-order non-Abelian group. As a result of its odd order, the corresponding representation category, ${\rm Rep}(\mathbb{Z}_7\rtimes\mathbb{Z}_3)$, does not have a braided charge conjugation symmetry \cite{MR3421083}.

\begin{center}
\begin{table}[h]
\centering
\begin{tabular}{c|ccccc}
& $C_e$ & $C_{a}$ & $C_{a^3}$ & $C_{b}$ & $C_{b^2}$ \\
\hline
$|C_x|$ & 1 & 3 & 3 & 7 & 7 \\
\hline
$\mathbf{1}$ & 1 & 1 & 1 & 1 & 1 \\
$\chi$ & 1 & 1 & 1 & $\zeta_3$ & $\zeta_3^2$ \\
$\bar\chi$ & 1 & 1 & 1 & $\zeta_3^2$ & $\zeta_3$ \\
$\rho$ & 3 & $\alpha$ & $\bar\alpha$ & 0 & 0 \\
$\overline\rho$ & 3 & $\bar\alpha$ & $\alpha$ & 0 & 0 \\
\end{tabular}
\caption{Character table for $\mathbb{Z}_7\rtimes\mathbb{Z}_3$, with $\zeta_3:=e^{2\pi i/3}$, $\zeta_7:=e^{2\pi i/7}$, and $\alpha:=\tfrac{-1+i\sqrt7}{2}=\zeta_7+\zeta_7^2+\zeta_7^4$ (therefore $\bar\alpha=\zeta_7^3+\zeta_7^5+\zeta_7^6$).}
\label{tab:char-Z7Z3}
\end{table}
\end{center}

To explicitly exhibit inherently complex $F$-symbols in this example, let us first define our group as
\begin{equation}
\mathbb{Z}_7\rtimes\mathbb{Z}_3:=\langle a,b|a^7=b^3=1,bab^{-1}=a^2\rangle~,
\end{equation}
where $\mathbb{Z}_3$ acts on $\mathbb{Z}_7$ via an order-three automorphism. Using this definition, any element can be written in the form $a^ib^j$ ($0\le i\le6$, $0\le j\le 2$) and lives in one of five conjugacy classes described in Table \ref{tab:char-Z7Z3}.

To understand the $F$-symbols and their gauge transformations, we need to study the five irreps and associated fusion rules listed in Table \ref{tab:fusion-Z7Z3}.  Given this data, let us examine the particularly simple $F$-symbol, $(F^{\chi\rho\rho}_{\rho})_{\rho\rho}$. First, note that it is one dimensional. More importantly, this $F$ symbol is also gauge invariant because
\begin{equation}\label{FINV}
(\widetilde{F}^{\chi\rho\rho}_{\rho})_{\rho\rho}=\Gamma^{\chi\rho}_{\rho}\Gamma^{\rho\rho}_{\rho} (F^{\chi\rho\rho}_{\rho})_{\rho\rho} (\Gamma^{\rho\rho}_{\rho})^{-1}(\Gamma^{\chi\rho}_{\rho})^{-1}=(F^{\chi\rho\rho}_{\rho})_{\rho\rho}~.
\end{equation}

\begin{table}[h]
\centering
\renewcommand{\arraystretch}{1.2}
\begin{tabular}{c|ccccc}
$\otimes$ & $\mathbf{1}$ & $\chi$ & $\bar\chi$ & $\rho$ & $\bar\rho$ \\
\hline
$\mathbf{1}$   & $\mathbf{1}$   & $\chi$        & $\bar\chi$    & $\rho$       & $\bar\rho$ \\
$\chi$         & $\chi$         & $\bar\chi$    & $\mathbf{1}$  & $\rho$       & $\bar\rho$ \\
$\bar\chi$     & $\bar\chi$     & $\mathbf{1}$  & $\chi$        & $\rho$       & $\bar\rho$ \\
$\rho$         & $\rho$         & $\rho$        & $\rho$        & $\rho\oplus 2\bar\rho$ & $\mathbf{1}\oplus\chi\oplus\bar\chi\oplus\rho\oplus\bar\rho$ \\
$\bar\rho$     & $\bar\rho$     & $\bar\rho$    & $\bar\rho$    & $\mathbf{1}\oplus\chi\oplus\bar\chi\oplus\rho\oplus\bar\rho$ & $2\rho\oplus\bar\rho$ \\
\end{tabular}
\caption{Fusion rules for the irreducible representations of $\mathbb{Z}_7\rtimes\mathbb{Z}_3$.}
\label{tab:fusion-Z7Z3}
\end{table}

To set up our computation of $F^{\chi\rho\rho}_{\rho}$, we first note that Table \ref{tab:char-Z7Z3} implies the following restriction
\begin{equation}
\rho|_N\cong \chi_1\oplus\chi_2\oplus\chi_4~,
\end{equation}
where $N:=\langle a\rangle\cong\mathbb{Z}_7$ is a normal subgroup, and $\chi_g$ is the irrep of $N$ satisfying $\chi_g(a)=\zeta_7^g$. Therefore, we can choose a basis $\left\{\psi_1,\psi_2,\psi_4\right\}$ for $V_{\rho}$ (the three-dimensional vector space associated with $\rho$) such that
\begin{equation}
\rho(a)\psi_k=\zeta_7^k\psi_k~, \ \ \ \rho(b)\psi_k=\psi_{4k \ {\rm mod}\ 7}~.
\end{equation}
The second equation above can be derived as follows: 
\begin{equation}
\rho(a)\rho(b)\psi_k = \rho(b)\rho(b^{-1}ab)\psi_k = \rho(b)\rho(a^{4})\psi_k = \zeta_7^{\,4k}\rho(b)\psi_k~.
\end{equation}

Next let us examine the fusion spaces and intertwiners that are involved in constructing $F^{\chi\rho\rho}_{\rho}$. The intertwiners we construct are in the fusion spaces ${\rm Hom}_G(V_{\chi}\otimes V_{\rho},V_{\rho})$ and ${\rm Hom}_G(V_{\rho}\otimes V_{\rho},V_{\rho})$. However, we compute the $F$-symbol using overlaps involving the corresponding adjoint splitting spaces via the prescription in Fig. \ref{fig:fr-moves}.

To that end, note that $\mathrm{Hom}_G(V_\chi\otimes V_\rho, V_\rho)$ is one-dimensional, and we can fix an isomorphism by choosing a unit vector $e_\chi\in V_\chi$. An intertwiner, $\varphi$, is represented by a linear map $M:V_\rho\to V_\rho$ determined by $\varphi(e_\chi\otimes\psi_k)=M\psi_k$, where $\varphi$ satisfies the equivariance condition $\varphi\circ(\chi(g)\otimes\rho(g))=\rho(g)\circ\varphi$. From Table \ref{tab:char-Z7Z3}, the equivariance condition reduces to commutativity on $g\in N$, so
$M\psi_k=m_k\psi_k$. Evaluating the equivariance condition on $e_\chi\otimes\psi_k$ at $g=b$ gives $\chi(b)\,m_{4k\bmod 7}=m_k$, so that $\chi(b)=\zeta_3$ implies the relation $m_{4k\bmod 7}=\zeta_3^{-1}m_k$. Normalizing $m_1=1$,\footnote{This normalization does not affect the final result for the $F$-symbol.} we conclude that
\begin{equation}
M = \mathrm{diag}(m_1,m_2,m_4)=\mathrm{diag}(1,\ \zeta_3,\ \zeta_3^{2})~,
\label{eq:M-Z7Z3}
\end{equation}
in the ordered basis $\left\{\psi_1,\psi_2,\psi_4\right\}$. Note that $M$ and $\varphi$ are unitary.

Finally, consider $\mathrm{Hom}_G(V_\rho\otimes V_\rho, V_\rho)$ and $f\in \mathrm{Hom}_G(V_\rho\otimes V_\rho, V_\rho)$. We write
$f(\psi_i\otimes\psi_j)=\sum_{k} \nu^{k}_{ij}\,\psi_k$ and impose
$f\circ(\rho(g)\otimes\rho(g))=\rho(g)\circ f$. Equivariance for $g=a$ implies
\begin{equation}
\nu^{k}_{ij}=0 \quad\text{unless}\quad i+j= k\ {\rm mod}\ 7~,
\label{eq:a-equiv-Z7Z3}
\end{equation}
with $i,j,k\in\{1,2,4\}$. Equivariance for $g=b$ results in
\begin{equation}
\nu^{\,4k\bmod 7}_{\,4i\bmod 7,\,4j\bmod 7} = \nu^{k}_{ij}~.
\label{eq:b-equiv-Z7Z3}
\end{equation}
Therefore, the matrix elements are invariant under the simultaneous action of $4\in\mathbb{Z}_7^\times$. The
admissible triples $(i,j,k)$ with $i,j,k\in\{1,2,4\}$ and $i+j= k$ mod 7 are
\begin{equation}
(1,1,2)~,\ \ \  (2,2,4)~,\ \ \ (4,4,1)~,
\label{eq:adm-Z7Z3}
\end{equation}
which form a single orbit
\begin{equation}
\CO_1 := \{(1,1,2),\,(4,4,1),\,(2,2,4)\}~.
\label{eq:orbit-Z7Z3}
\end{equation}
Hence there is one complex degree of freedom remaining, which agrees with the fusion rules of Table~\ref{tab:fusion-Z7Z3} (i.e., $N^{\rho}_{\rho\rho}:=\dim\mathrm{Hom}_G(V_\rho\otimes V_\rho,V_\rho)=1$). We define the (single)
orbit-basis intertwiner
\begin{equation}
\nu(\psi_i\otimes\psi_j)=\sum_{k\,:\,(i,j,k)\in \CO_1}\psi_k=
\begin{cases}
\psi_{i+j} & (i,j,i+j)\in \CO_1~,\\
0 & \text{otherwise~,}
\end{cases}
\label{eq:nu-Z7Z3}
\end{equation}
which satisfies $\nu\nu^\dagger=\mathrm{Id}_{V_\rho}$.

We can now assemble our ingredients and compute the $F$-symbol as the normalized overlap of the two associativity paths (the two sides of the top diagram in Fig. \ref{fig:fr-moves})\footnote{Note that $\varphi$ and $\nu$ were defined for fusion spaces, whereas Fig. \ref{fig:fr-moves} is stated in terms of splitting spaces. Therefore Fig. \ref{fig:fr-moves} translates directly into an inner product involving $\varphi^\dagger$ and $\nu^\dagger$ (if we so wish, we can rewrite the inner product in terms of $\varphi$ and $\nu$ using antiunitarity of the dagger). Had we instead taken the $F$-symbol to be defined in terms of fusion spaces, we would obtain the complex conjugate of our $F$-symbol. Either convention exhibits the same inherent complexity.\label{SplitFus}}
\begin{eqnarray}
\bigl(F^{\chi\rho\rho}_\rho\bigr)_{\rho\rho}
&=& \bigl\langle\, (\mathrm{id}_{V_\chi}\otimes \nu^{\dagger})\circ\varphi^{\dagger} \,\big|\,
  (\varphi^\dagger\otimes \mathrm{id}_{V_\rho})\circ\nu^\dagger\,\bigr\rangle\cr&=&\frac{1}{\dim V_\rho}\,\mathrm{Tr}_{V_\chi\otimes V_\rho\otimes V_\rho}\!
  \Bigl[\bigl(\nu\circ(\varphi\otimes\mathrm{id}_{V_\rho})\bigr)^\dagger\circ\varphi\circ(\mathrm{id}_{V_\chi}\otimes\nu)\Bigr]~,
  \label{eq:F-def-Z7Z3}
\end{eqnarray}
where we have used the definition of the bracket and cyclicity of the trace. Defining
\begin{equation}
L := \nu\circ(\varphi\otimes \mathrm{id}_{V_\rho})~,\ \ \ 
R := \varphi\circ(\mathrm{id}_{V_\chi}\otimes \nu)~,
\label{eq:LR-Z7Z3}
\end{equation}
we evaluate
\begin{align}
L(e_\chi\otimes\psi_i\otimes\psi_j)
&=\nu(m_i\psi_i\otimes\psi_j)=
\begin{cases}m_i\,\psi_{i+j} & (i,j,i+j)\in \CO_1~,\\ 0&\text{otherwise~,}\end{cases}
\label{eq:L-eval-Z7Z3}\\[4pt]
R(e_\chi\otimes\psi_i\otimes\psi_j)
&=\varphi\!\left(e_\chi\otimes \nu(\psi_i\otimes\psi_j)\right)=
\begin{cases}m_{i+j}\,\psi_{i+j} & (i,j,i+j)\in \CO_1,\\ 0&\text{otherwise~.}\end{cases}
\label{eq:R-eval-Z7Z3}
\end{align}
Both $L$ and $R$ have exactly three nonzero matrix elements (one for each element in $\CO_1$).

Plugging into \eqref{eq:F-def-Z7Z3}, we obtain our promised measure of inherent complexity
\begin{equation}
\bigl(F^{\chi\rho\rho}_\rho\bigr)_{\rho\rho}
= \frac{1}{3}\sum_{(i,j,k)\in \CO_1}\overline{m_i}\,m_{k}
= \overline{m_i}\,m_{i+j}\big|_{\CO_1}
=\zeta_3~,
\label{eq:F-result-Z7Z3}
\end{equation}
where the second equality holds because $\overline{m_i}\,m_{i+j}$ is constant along $\CO_1$. This result is complex and gauge invariant as claimed.

\subsec{\texorpdfstring{${\rm Rep}(\mathbb{Z}_5\rtimes\mathbb{Z}_4)$}{Rep(Z5 semidirect product Z4)}}
\label{Z5Z4}
As our second example, we consider the smallest group, $\mathbb{Z}_5\rtimes Z_4$, whose representation category has inherently complex $F$-symbols. By our above discussion, this complexity carries over to all corresponding discrete gauge theories based on the group. Our computation is similar to the one in the previous section, but here our statement will be (by necessity) about a combination of $F$-symbols. As in the previous example, our reason for studying this group is that it lacks a class-inverting automorphism and hence does not have a braided charge-conjugation symmetry acting on its representation category \cite{MR3421083}.

The group we wish to study has the following presentation
\begin{equation}
\mathbb{Z}_5\rtimes\mathbb{Z}_4:=\langle a,b|a^5=b^4=1,bab^{-1}=a^2\rangle~,
\end{equation}
where $\mathbb{Z}_4$ acts on $\mathbb{Z}_5$ via an order-four automorphism. Using the above definition, any element can be written in the form $a^ib^j$ ($0\le i\le4$, $0\le j\le 3$) and lives in one of five conjugacy classes described in Table \ref{tab:char-F20}.

\begin{table}[h]
\centering
\begin{tabular}{c|ccccc}
 & $C_e$ & $C_a$ & $C_{b}$ & $C_{b^3}$ & $C_{b^2}$ \\
\hline
$|C_{x}|$ & 1 & 4 & 5 & 5 & 5 \\
\hline
$\mathbf{1}$ & 1 & 1 & 1 & 1 & 1 \\
$\chi$ & 1 & 1 & $i$ & $-i$ & $-1$ \\
$\varepsilon$ & 1 & 1 & $-1$ & $-1$ & 1 \\
$\chi^{*}$ & 1 & 1 & $-i$ & $i$ & $-1$ \\
$\rho$ & 4 & $-1$ & 0 & 0 & 0 \\
\end{tabular}
\caption{Character table for $\mathbb{Z}_5\rtimes\mathbb{Z}_4$.}
\label{tab:char-F20}
\end{table}

Let us first show that this group lacks a class-inverting automorphism, $\varphi$. To that end, such a $\varphi$ would have to map $\varphi(C_e)=C_e$, $\varphi(C_a)=C_a$, $\varphi(C_b)=C_{b^3}$, and $\varphi(C_{b^3})=C_b$. As a result, we would need to have $\varphi(a)=a^k$ and $\varphi(b)=a^{\ell}b^3$ for some $k$ and $\ell$. But then the defining relation of the group becomes
\begin{equation}
\varphi(bab^{-1})=a^{\ell}b^3a^kb^{-3}a^{-\ell}=a^{3k}=\varphi(a^2)=a^{2k}~.
\end{equation}
However, this identification implies $a^k=1$, which is a contradiction.

\begin{table}[h]
\centering
\renewcommand{\arraystretch}{1.2}
\begin{tabular}{c|ccccc}
$\otimes$ & $\mathbf{1}$ & $\chi$ & $\varepsilon$ & $\chi^{*}$ & $\rho$ \\
\hline
$\mathbf{1}$  & $\mathbf{1}$  & $\chi$         & $\varepsilon$ & $\chi^{*}$    & $\rho$ \\
$\chi$        & $\chi$        & $\varepsilon$  & $\chi^{*}$    & $\mathbf{1}$  & $\rho$ \\
$\varepsilon$ & $\varepsilon$ & $\chi^{*}$     & $\mathbf{1}$  & $\chi$        & $\rho$ \\
$\chi^{*}$    & $\chi^{*}$    & $\mathbf{1}$   & $\chi$        & $\varepsilon$ & $\rho$ \\
$\rho$        & $\rho$        & $\rho$         & $\rho$        & $\rho$        & $\mathbf{1} \oplus \chi \oplus \varepsilon \oplus \chi^{*} \oplus 3\rho$ \\
\end{tabular}
\caption{Fusion rules for the irreducible representations of $\mathbb{Z}_5\rtimes\mathbb{Z}_4$.}
\label{tab:fusion-F20}
\end{table}

Therefore, we are motivated to check for complex $F$. To that end, consider
\begin{equation}\label{Fz4z5}
F^{\chi\rho\rho}_{\rho}:=\left(F^{\chi\rho\rho}_{\rho}\right)_{(\rho,\alpha),(\rho,\beta)}~,\ \ \ F^{\rho\rho\chi}_{\rho}:=\left(F^{\rho\rho\chi}_{\rho}\right)_{(\rho,\alpha),(\rho,\beta)}~,
\end{equation}
with $1\le\alpha,\beta\le3$. We claim that 
\begin{equation}\label{Fcomplex}
F:=F^{\chi\rho\rho}_{\rho}\left(F^{\rho\rho\chi}_{\rho}\right)^{-1}~,
\end{equation}
is inherently complex (even though there are gauges in which one of the $F$-symbols in \eqref{Fz4z5} is real). In particular, the gauge-invariant spectrum of the resulting matrix is
\begin{equation}
{\rm spec}\left(F\right)=\left\{-1,i,i\right\}~.
\end{equation}
Therefore, $F$ is complex in any gauge. For example, one simple gauge-invariant detector of this inherent complexity is
\begin{equation}
{\rm Tr}(F)\ne{\rm Tr}(\overline{F})~.
\end{equation}

To understand the above discussion more quantitatively, consider the fusion rules of ${\rm Rep}(\mathbb{Z}_5\rtimes\mathbb{Z}_4)$ in Table \ref{tab:fusion-F20}. From this data, we can derive the following gauge transformation properties for $F^{\chi\rho\rho}_{\rho}$
\begin{eqnarray}
\left(\widetilde{F}^{\chi\rho\rho}_{\rho}\right)_{(\rho,\alpha),(\rho\beta)}&=&\sum_{\mu,\nu}\Gamma^{\chi\rho}_{\rho}\left(\Gamma^{\rho\rho}_{\rho}\right)_{\alpha,\mu}\left(F^{\chi\rho\rho}_{\rho}\right)_{(\rho,\mu),(\rho,\nu)}\left(\Gamma^{\rho\rho}_{\rho}\right)_{\nu,\beta}^{-1}\left(\Gamma^{\chi\rho}_{\rho}\right)^{-1}\cr&=&\sum_{\mu,\nu}\left(\Gamma^{\rho\rho}_{\rho}\right)_{\alpha,\mu}\left(F^{\chi\rho\rho}_{\rho}\right)_{(\rho,\mu),(\rho,\nu)}\left(\Gamma^{\rho\rho}_{\rho}\right)_{\nu,\beta}^{-1}~,
\end{eqnarray}
and $F^{\rho\rho\chi}_{\rho}$ 
\begin{eqnarray}
\left(\widetilde{F}^{\rho\rho\chi}_{\rho}\right)_{(\rho,\alpha),(\rho\beta)}&=&\sum_{\mu,\nu}\left(\Gamma^{\rho\rho}_{\rho}\right)_{\alpha,\mu}\Gamma^{\rho\chi}_{\rho}\left(F^{\rho\rho\chi}_{\rho}\right)_{(\rho,\mu),(\rho,\nu)}\left(\Gamma^{\rho\chi}_{\rho}\right)^{-1}\left(\Gamma^{\rho\rho}_{\rho}\right)_{\nu,\beta}^{-1}\cr&=&\sum_{\mu,\nu}\left(\Gamma^{\rho\rho}_{\rho}\right)_{\alpha,\mu}\left(F^{\rho\rho\chi}_{\rho}\right)_{(\rho,\mu),(\rho,\nu)}\left(\Gamma^{\rho\rho}_{\rho}\right)_{\nu,\beta}^{-1}~.
\end{eqnarray}
Therefore, we have the following transformation for $F$ in \eqref{Fcomplex}
\begin{eqnarray}
\widetilde{F}_{(\rho,\alpha),(\rho\beta)}&=&\sum_{\mu,\nu}\left(\Gamma^{\rho\rho}_{\rho}\right)_{\alpha,\mu}F_{(\rho,\mu),(\rho,\nu)}\left(\Gamma^{\rho\rho}_{\rho}\right)_{\nu,\beta}^{-1}~,
\end{eqnarray}
and so ${\rm spec}(F)$ and ${\rm Tr}(F)$ are indeed gauge invariant.

Ultimately, we will write $F^{\chi\rho\rho}_{\rho}$, $F^{\rho\rho\chi}_{\rho}$, and $F$ as diagonal matrices in the same basis. To arrive at this result, we first observe that, from Table \ref{tab:char-F20}, we have the following restriction of the non-Abelian representation
\begin{equation}
\rho|_N\cong\chi_1\oplus\chi_2\oplus\chi_3\oplus\chi_4~,
\end{equation}
where $N:=\langle a\rangle\cong\mathbb{Z}_5<\mathbb{Z}_5\rtimes\mathbb{Z}_4$ is a normal subgroup, and the $\chi_i$ are the four non-trivial irreps of $N$. Therefore, we can choose a basis $\left\{\psi_1,\psi_2,\psi_3,\psi_4\right\}$ of $V_{\rho}$ (the vector space associated with $\rho$) such that 
\begin{equation}
\rho(a)\psi_k=\zeta^k\psi_k~, \ \ \ \rho(b)\psi_k=\psi_{3k\ {\rm mod}\ 5}~,\ \ \ \zeta:=\exp(2\pi i/5)~.
\end{equation}
The second equality follows from the fact that $\rho(a)\rho(b)\psi_k=\rho(b)\rho(b^{-1}ab)\psi_k=\rho(b)\rho(a^3)\psi_k=\zeta^{3k}\rho(b)\psi_k$.

Next let us examine the fusion spaces and intertwiners that are involved in constructing the above $F$-symbols. The intertwiners we construct are in the fusion spaces ${\rm Hom}_G(V_{\chi}\otimes V_{\rho},V_{\rho})$, ${\rm Hom}_G(V_{\rho}\otimes V_{\chi},V_{\rho})$, and ${\rm Hom}_G(V_{\rho}\otimes V_{\rho},V_{\rho})$. However, we compute the $F$-symbol using overlaps involving the corresponding adjoint splitting spaces via the prescription in Fig. \ref{fig:fr-moves}.

We will consider the fusion spaces one-by-one. Note that ${\rm Hom}_G(V_{\chi}\otimes V_{\rho},V_{\rho})$ is one-dimensional and we can fix an isomorphism by choosing a unit vector $e_{\chi}\in V_{\chi}$. An intertwiner then becomes a linear map $M: V_{\rho}\to V_{\rho}$ determined by $\varphi(e_{\chi}\otimes \psi_k)=M\psi_k$, where $\varphi$ satisfies the equivariance condition, $\varphi\circ(\chi(g)\otimes\rho(g))=\rho(g)\circ\varphi$. In particular, from Table \ref{tab:char-F20} we see that the equivariance condition reduces to commutativity on $g\in N$. Therefore, $M\psi_k=m_k\psi_k$. Next, the equivariance condition combined with $\chi(b)=i$ implies that $m_k=im_{3k\ {\rm mod}\ 5}$. Normalizing $m_1=1$,\footnote{This normalization is arbitrary and does not affect the final results for our $F$-symbols.} we conclude that
\begin{equation}
M={\rm diag}(1,i,-i,-1)~,
\end{equation}
where $M$ and $\varphi$ are unitary.

Next let us consider ${\rm Hom}_G(V_{\rho}\otimes V_{\chi},V_{\rho})$. We can repeat the derivation in the previous paragraph almost verbatim. The corresponding intertwiner, $\varphi'$, becomes a linear map $M':V_{\rho}\to V_{\rho}$ determined by $\varphi'(\psi_k\otimes e_{\chi})=M'\psi_k$, where $\varphi'$ satisfies the equivariance condition, $\varphi'\circ(\rho(g)\otimes\chi(g))=\rho(g)\circ\varphi'$. By logic analogous to that in the previous paragraph, $M'\psi_k=m'_k\psi_k$, and  normalizing such that $m'_1=1$,\footnote{As in the case of $M$, this normalization is arbitrary and does not affect the final results of this section. Note also that, although we have chosen bases such that $M=M'$, $\varphi$ and $\varphi'$ are independent intertwiners with separate $U(1)$ gauge freedom.} we find
\begin{equation}
M'={\rm diag}(1,i,-i,-1)~.
\end{equation}
Note that both $M'$ and $\varphi'$ are unitary.

Finally, let us consider ${\rm Hom}_G(V_{\rho}\otimes V_{\rho},V_{\rho})$ and $f\in{\rm Hom}_G(V_{\rho}\otimes V_{\rho},V_{\rho})$. We define $f(\psi_i\otimes\psi_j)=\sum_{k=1}^4\nu_{ij}^k\psi_k$ and impose the equivariance condition, $f\circ(\rho(g)\otimes\rho(g))=\rho(g)\circ f$. The equivariance condition for $g=a$ forces
\begin{equation}
\nu_{ij}^k=0~, \ \ \ {\rm unless}\ \ \ i+j=k\ {\rm mod}\ 5~.
\end{equation}
This constraint leaves over twelve degrees of freedom. On the other hand, equivariance for $g=b$ forces
\begin{equation}
\nu_{3i \ {\rm mod}\ 5, 3j \ {\rm mod}\ 5}^{3k \ {\rm mod}\ 5}=\nu_{ij}^k~.
\end{equation}
In particular, the $\nu_{ij}^k$ matrix elements are constant under the action of $3\in\mathbb{Z}_5^{\times}$. Indexing the matrix elements by $(i,j,k)$ triples, we then have the following three orbits
\begin{eqnarray}\label{orbits}
\CO_1&=&\left\{(1,1,2),(3,3,1),(4,4,3),(2,2,4)\right\}~, \cr \CO_2&=&\left\{(1,2,3),(3,1,4),(4,3,2),(2,4,1)\right\}~, \cr \CO_3&=&\left\{(1,3,4),(3,4,2),(4,2,1),(2,1,3)\right\}~.
\end{eqnarray}
As a result, we have three complex degrees of freedom left over, which is what we expect from the fusion rules in Table \ref{tab:fusion-F20} (i.e., $N_{\rho\rho}^{\rho}:={\rm dim}\left({\rm Hom}_G(V_{\rho}\otimes V_{\rho},V_{\rho})\right)=3$). It is natural to now define an orbit basis for this hom-space via
\begin{equation}
\nu_r(\psi_i \otimes \psi_j) = \sum_{k\ {\rm s.t.}\ (i,j,k)\in \CO_r}\psi_k =
\begin{cases}
\psi_{i+j} & \text{if } (i,j,i+j)\in \CO_r~,\\[2pt]
0 & \text{otherwise}~,
\end{cases}
\end{equation} 
for $r=1,2,3$. It is easy to check that 
\begin{equation}\label{nuOrthog}
\nu_r\nu_s^{\dagger}=\delta_{rs}{\rm Id}_{V_{\rho}}~.
\end{equation}

We now have all the ingredients to compute the $F$-symbols. Indeed, using logic similar to that in the previous example, we have (daggers of our intertwiners appear for the same reason as in Footnote \ref{SplitFus})
\begin{eqnarray}\label{Fcrr}
\left(F^{\chi\rho\rho}_{\rho}\right)_{(\rho,\mu),(\rho,\nu)}&=&\langle ({\rm id}_{V_{\chi}}\otimes \nu_{\nu}^\dagger)\circ\varphi^\dagger|(\varphi^\dagger\otimes {\rm id}_{V_{\rho}})\circ\nu_{\mu}^\dagger\rangle
\cr&=&{1\over\dim V_\rho}{\rm Tr}_{V_{\chi}\otimes V_{\rho}\otimes V_{\rho}}\left(\left(\nu_{\mu}\circ(\varphi\otimes {\rm id}_{V_{\rho}})\right)^{\dagger}\circ\varphi\circ({\rm id}_{V_{\chi}}\otimes \nu_{\nu})\right)~,\ \ \ \ \ \ 
\end{eqnarray}
where we have used the definition of the bracket and cyclicity of the trace. 

To find the $F$-symbol, we need to compute the action of
\begin{equation}
L_{\mu}:=\nu_{\mu}\circ(\varphi\otimes{\rm id}_{V_{\rho}})~,\ \ \ R_{\nu}:=\varphi\circ({\rm id}_{V_{\chi}}\otimes\nu_{\nu})~,
\end{equation}
on $V_{\chi}\otimes V_{\rho}\otimes V_{\rho}$. In particular, we find
\begin{eqnarray}
L_{\mu}(e_{\chi}\otimes\psi_i\otimes\psi_j)&=&\nu_{\mu}(\varphi(e_{\chi}\otimes\psi_i)\otimes\psi_j)=\nu_{\mu}(m_i\psi_i\otimes\psi_j)=m_i\nu_{\mu}(\psi_i\otimes\psi_j)\cr&=&\begin{cases}
m_i\psi_{i+j} & \text{if } (i,j,i+j)\in \CO_{\mu}~,\\[2pt]
0 & \text{otherwise}~.
\end{cases}
\end{eqnarray}
Similarly, we have
\begin{eqnarray}
R_{\nu}(e_{\chi}\otimes\psi_i\otimes\psi_j)&=&\varphi\left(e_{\chi}\otimes\nu_{\nu}(\psi_i\otimes\psi_j)\right)=\varphi\left(e_{\chi}\otimes\sum_{k\ {\rm s.t.}\ (i,j,k)\in \CO_{\nu}}\psi_k\right)\cr&=&\begin{cases}
m_{i+j}\psi_{i+j} & \text{if } (i,j,i+j)\in \CO_{\nu}~,\\[2pt]
0 & \text{otherwise}~.
\end{cases}
\end{eqnarray}
Note that both $L_{\mu}$ and $R_{\nu}$ have, for fixed $\mu$ and $\nu$, four non-zero matrix elements, $(L_{\mu})_{ij,i+j}$ and $(R_{\nu})_{ij,i+j}$ (corresponding to the elements of the orbits in \eqref{orbits}). Finally, plugging into \eqref{Fcrr}, we have 
\begin{eqnarray}\label{Fcrr2}
\left(F^{\chi\rho\rho}_{\rho}\right)_{(\rho,\mu),(\rho,\nu)}&=&{1\over4}\sum_{i,j,k}(\overline{L}_{\mu})_{ij,k}(R_{\nu})_{ij,k}={1\over4}\delta_{\mu\nu}\sum_{(i,j,k)\in\CO_{\mu}}\overline{m}_i m_{i+j}={\rm diag}(i,-i,-1)~.\ \ \ 
\end{eqnarray}

Next, let us compute $F^{\rho\rho\chi}_{\rho}$. By similar logic, we have (again the same discussion as in Footnote \ref{SplitFus} applies)
\begin{eqnarray}\label{Frrc}
\left(F^{\rho\rho\chi}_{\rho}\right)_{(\rho,\mu),(\rho,\nu)}&=&\langle ({\rm id}_{V_{\rho}}\otimes \varphi'^\dagger)\circ\nu_{\nu}^\dagger|(\nu_{\mu}^\dagger\otimes {\rm id}_{V_{\chi}})\circ \varphi'^\dagger\rangle
\cr&=&{1\over\dim V_\rho}{\rm Tr}_{V_{\rho}\otimes V_{\rho}\otimes V_{\chi}}\left(\left(\varphi'\circ(\nu_{\mu}\otimes {\rm id}_{V_{\chi}})\right)^{\dagger}\circ\nu_{\nu}\circ({\rm id}_{V_{\rho}}\otimes \varphi')\right)~,\ \ \ \ \ \ \ \ 
\end{eqnarray}
where we have used the definition of the bracket and cyclicity of the trace. As before, we define
\begin{equation}
L_{\mu}':=\varphi'\circ(\nu_{\mu}\otimes{\rm id}_{V_{\chi}})~,\ \ \ R'_{\nu}:=\nu_{\nu}\circ({\rm id}_{V_{\rho}}\otimes\varphi')~,
\end{equation}
and compute
\begin{eqnarray}
L'_{\mu}(\psi_i\otimes\psi_j\otimes e_{\chi})&=&\varphi'(\nu_{\mu}(\psi_i\otimes\psi_j)\otimes e_{\chi})=\varphi'\left(\sum_{k\ {\rm s.t.}\ (i,j,k)\in \CO_{\mu}}\psi_k\otimes e_{\chi}\right)\cr&=&\begin{cases}
m_{i+j}\psi_{i+j} & \text{if } (i,j,i+j)\in \CO_{\mu}~,\\[2pt]
0 & \text{otherwise}~.
\end{cases}
\end{eqnarray}
Similarly, we have
\begin{eqnarray}
R'_{\nu}(\psi_i\otimes\psi_j\otimes e_{\chi})&=&\nu_{\nu}(\psi_i\otimes\varphi'(\psi_j\otimes e_{\chi}))=\nu_{\nu}\left(\psi_i\otimes m_j\psi_j\right)\cr&=&\begin{cases}
m_{j}\psi_{i+j} & \text{if } (i,j,i+j)\in \CO_{\nu}~,\\[2pt]
0 & \text{otherwise}~.
\end{cases}
\end{eqnarray}
Finally, plugging into \eqref{Frrc}, we have 
\begin{eqnarray}\label{Frrc2}
\left(F^{\rho\rho\chi}_{\rho}\right)_{(\rho,\mu),(\rho,\nu)}&=&{1\over4}\sum_{i,j,k}(\overline{L}'_{\mu})_{ij,k}(R'_{\nu})_{ij,k}={1\over4}\delta_{\mu\nu}\sum_{(i,j,k)\in\CO_{\mu}}\overline{m}_{i+j}m_{j}={\rm diag}(-i,-1,i)~.\ \ \ 
\end{eqnarray}
Therefore, the ratio of $F$-symbols in \eqref{Fcomplex} satisfies
\begin{eqnarray}\label{Ffinal}
F&=&{\rm diag}(-1,i,i)~,
\end{eqnarray}
which can never be made real since
\begin{equation}\label{complexTrace}
{\rm Tr}(F)=-1+2i\ne-1-2i={\rm Tr}(\overline F)~.
\end{equation}
This is our promised measure of inherent complexity. As a useful check of this result, we note that \eqref{complexTrace} is consistent with the $F$-symbols derived in Example 3 of \cite{siehler2003near}.\footnote{To understand this relation with \cite{siehler2003near} quantitatively, the interested reader should use the following dictionary: $\chi\leftrightarrow g$, $\varepsilon\leftrightarrow g^2$, $\chi^*\leftrightarrow g^3$, and $\rho\leftrightarrow m$. Then, up to a gauge transformation, we have $F^{\chi\rho\rho}_{\rho}\leftrightarrow\gamma_1(g)$ and $F^{\rho\rho\chi}_{\rho}\leftrightarrow\gamma_3(g)$. In particular, we have the gauge-invariant matching ${\rm Tr}(F)={\rm Tr}(\gamma_1(g)\gamma_3(g)^{-1})$. The curious reader can check that the gauge transformation itself is trivial except on ${\rm Hom}(\rho\otimes\rho,\rho)$, where it is
\begin{equation}
\Gamma^{\rho\rho}_{\rho}=\begin{pmatrix}
1&0&0\\
0&0&1\\
0&1&0
\end{pmatrix}~.
\end{equation}
}

\section{Conclusion and Discussion}

In this paper, we have explored a converse direction to the result in \cite{ToAppear}. In particular, we have exhibited inherently complex $F$-symbols in the following premodular (braided) fusion categories that lack a braided charge-conjugation autoequivalence: ${\rm Rep}(\mathbb{Z}_7\rtimes\mathbb{Z}_3)$ and ${\rm Rep}(\mathbb{Z}_5\rtimes\mathbb{Z}_4)$, along with their Drinfeld centers. Our work leaves many interesting open questions. Among them:

\begin{itemize}
\item We have hinted at a potential connection between inherently complex $F$-symbols and UMTCs not determined by their modular data \cite{mignard2021modular}. It would be interesting to understand the precise relation.

\item Are there any experimental realizations of, or uses for, the complex phases we found in Sec. \ref{Z7Z3} and Sec. \ref{Z5Z4}? More abstractly, do the complex phases we find have interesting consequences at the level of Galois theory, or for the Galois action on topological data (e.g., see \cite{Buican:2021axn})?

\item Our main examples were all unitary. What happens when one relaxes unitarity? For example, as in the case of Fibonacci anyons, Lee-Yang has a gauge with real $F$-symbols.

\item We constructed tailored gauge-invariant detectors of $F$-symbol complexity when studying our examples. Are there useful and universal gauge-invariant detectors of $F$-symbol inherent complexity? For example, the observables in \cite{Lake:2016mky} do not detect inherent complexity in our examples.

\item It would be interesting to work out the inherent complexity of the $F$-symbols in the generalized families ${\rm Rep}(\mathbb{Z}_p\rtimes\mathbb{Z}_q)$ and ${\rm Rep}(\mathbb{F}_{q'}\rtimes\mathbb{F}_{q'}^{\times})$ that include our categories as the simplest cases.

\item It would be interesting to determine whether there are smaller-rank unitary premodular categories with inherently complex $F$-symbols.
\end{itemize}

We hope to return to some of these questions soon.

\ack{We thank D.~Nikshych, K.~Walker, and Z.~Wang for discussions. We are grateful to the Isaac Newton Institute for Mathematical Sciences at Cambridge University for support and hospitality during the programme \lq\lq Quantum Field Theory With Boundaries, Impurities, and Defects,” where this project was initiated. We were partly supported by EPSRC grant no. EP/Z000580/1. M. B.'s work was partly supported by STFC under the grant “Amplitudes, Strings and Duality” and by a Simons Fellowship. {P.H.}~was supported by EPSRC Programme Grant {No.}~EP/W007509/1. J.K.P. was partially supported by EPSRC grants nos. EP/W524372/1 and UKRI1337:Anyons24. No new data were generated or analyzed in this study.}

\newpage
\bibliography{chetdocbib}

\end{document}